%% file: task-summa.tex
\documentclass{sig-alternate}

\usepackage{amsmath}
\usepackage{mathtools}		
\usepackage{fixltx2e}		
\usepackage{graphicx}		
\usepackage{subcaption}
\usepackage{bm} 			
\usepackage{algpseudocode}	
\usepackage{algorithm}		
\usepackage{siunitx}                  
\usepackage{tikz}
\usetikzlibrary{positioning,shapes.geometric}

\def\sharedaffiliation{%
\end{tabular}
\begin{tabular}{c}}

\usepackage[colorinlistoftodos]{todonotes}

\begin{document}
%
\conferenceinfo{{\it IA}$^3$ 2015}{November 15-20, 2015, Austin, TX, USA}
\CopyrightYear{2015} 
\crdata{ISBN 978-1-4503-4001-4/15/11}  
\DOI{10.1145/2833179.2833186}

\title{Scalable Task-Based Algorithm for Multiplication of Block-Rank-Sparse Matrices}
%
%
%
%
%

\numberofauthors{3} 
%
\author{
%
%
  \alignauthor Justus A. Calvin\\   
  \email{justusc@vt.edu} \\
  \alignauthor Cannada A. Lewis\\     
  \email{calewis@vt.edu} \\
  \alignauthor Edward F. Valeev*\\     
  \email{efv@vt.edu}
  \sharedaffiliation
  \affaddr{Department of Chemistry}  \\
  \affaddr{Virginia Tech}\\
  \affaddr{Blacksburg, VA}\\
}

\maketitle

\begin{abstract}
%
%

%
A task-based formulation of Scalable Universal Matrix Multiplication Algorithm (SUMMA), a popular algorithm for matrix multiplication (MM), is applied to the multiplication of hierarchy-free, rank-structured matrices that appear in the domain of quantum chemistry (QC).
The novel features of our formulation are: (1) concurrent scheduling of multiple SUMMA iterations, and (2) fine-grained task-based composition. These features make it tolerant of the load imbalance due
to the irregular matrix structure and 
eliminate all artifactual sources of global synchronization. 
Scalability of iterative computation of square-root inverse of block-rank-sparse QC matrices is demonstrated; for full-rank (dense) matrices the performance of our SUMMA formulation usually exceeds that of the state-of-the-art dense MM implementations (ScaLAPACK and Cyclops Tensor Framework).

\end{abstract}





\section{Introduction}


A frontier challenge posed by scientific and engineering applications in areas as distinct as quantum physics and machine learning is dealing with sparse and non-standard tensorial data representations. Such data appears in many forms: sparse tensors, multiresolution spectral element trees, $\mathcal{H}$-matrices, tensor networks, and many others. What they share in common
is the reduced number of parameters in terms of which the data is represented, at the cost of more complex data representation and computation relative to the standard/na\"ive counterpart. The need to deal with irregular data representations conflicts with the evolution of computer hardware and programming models that demand regular patterns of data access and computation for peak performance. This tension drives the search for algorithms that minimize/avoid communication and/or hide its cost, and are capable of dealing with increasingly irregular numerical data structures.

In this work we explore parallel computation with matrices composed of low-rank blocks that we refer to as Clustered Low-Rank  (CLR) matrices\footnote{Related matrix data structures have appeared under many names (matrices with decay, $\mathcal{H}$-matrices, rank-structured matrices, and mosaic skeleton approximation), but no single globally-accepted terminology exists. For the history of these types of matrices see Ref \cite{Vandebril:2005fi}.}
Concrete examples of such matrices are taken from quantum chemistry and materials science. Matrices (tensors) in such context represent quantum states (of electrons) and operators represented in some basis. Efficient application of operators to states --- represented by matrix multiplication (tensor contraction) --- demands taking advantage of the matrix (tensor) structure that take the form of (a) block-sparsity due to the distance decay of the operator kernel and the localized nature of basis functions \cite{Hollman:2015ca}, (b) symmetries under geometric and other transformations \cite{Rajbhandari:2014bx,Hirata:2003is}, and/or (c) block-rank-sparsity due to smoothness of states and operator kernels \cite{Neese:2009dz}. Notably, exploiting this matrix structure depends on problem-specific blocking of matrix dimensions that arise due to domain-specific needs and typically cannot be chosen arbitrarily.
In other words, the matrices that we encounter are ``sparse'' in a general sense, which encompasses element-, block-, and block-rank-sparsity; but in a practical sense the matrices are not sparse enough to be a good match for the established sparse MM algorithms.

The key idea is a novel task-based variant of Scalable Universal Matrix Multiplication Algorithm (SUMMA) of van de Geijn and Watts,\cite{VanDeGeijn:1997uf}, a well-known algorithm for
distributed-memory dense matrix multiplication (MM).
The nonuniform blocking and data inhomogeneity of CLR matrices conflict with the {\em uniform} data distribution exploited by all distributed-memory, dense MM algorithms --- including Cannon's \cite{Cannon:1969vd}, SUMMA \cite{VanDeGeijn:1997uf}, and others \cite{Fox:1987ve,Choi:1994jw,Buluc:2008vn,Solomonik:2014ko,Rajbhandari:2014bx}. Task-based formulation allows us to overcome this limitation.
Task-based/dataflow programming models are a natural choice for implementation of algorithms with irregular data and computation patterns; such models have already been used successfully for dense matrix algebra applications \cite{Husbands:2007de,pdsec}.
Besides handling matrices with structure, the task-based approach provides
additional benefits: (a) inter-node communication costs can be partially or fully hidden by overlapping computation and communication, (b) performance should be less sensitive to topology, latency, and CPU clock variations, (c) fine-grained, task-based parallelism is a proven means to attain high intra-node performance by leveraging massively multicore platforms and hiding the costs of memory hierarchy ({\em e.g.}\ Intel TBB, Cilk), (d) lack of global synchronization allows the overlap multiple, high-level stages of computation ({\em e.g.}\ two or more multiple matrix multiplications contributing to the same expression).


The new formulation was used to implement iterative computation of the square root inverse of a matrix,
a prototypical operation in which block ranks of intermediate matrices change dynamically
during the iteration. The usual advantage of the task formulation, tolerance of load imbalance and latency, are demonstrated in the regime where matrices approach full rank, by comparison against
the state-of-the-art dense MM implementations. 


\section{Standard SUMMA and Variants}

Before describing our algorithm, we start out by recapping the standard SUMMA and its variants for distributed-memory MM. Although there are earlier \cite{Cannon:1969vd} and asymptotically faster \cite{Strassen:1969im}
algorithms, SUMMA has become popular due to its relative simplicity and flexibility (it can be easily generalized to rectangular matrices and process grids). From a just as important practical standpoint, SUMMA,
like all 2D algorithms, uses memory more economically than its 3D counterparts \cite{Berntsen:1989bx,Agarwal:1995jl}.
SUMMA is also a building block for more complicated 2.5D and 3D MM algorithms \cite{Solomonik:2011bra}
and produced a number of variants,\cite{Choi:1997cp,Gunnels:1998gg} including sparse SUMMA (SpSUMMA) \cite{Buluc:2008vn,Buluc:2012bp}.

SUMMA implements matrix multiplication {\bf C} = {\bf A}{\bf B} as a series of rank-$k$ updates.
The input and output matrices are embedded on a rectangular process mesh in an element-cyclic or block-cyclic manner to ensure approximate load balance.
In each iteration of the algorithm, a column/row panel of {\bf A}/{\bf B} is broadcast along rows/columns of the process grid, respectively; matrix {\bf C} remains stationary throughout the procedure (variants of SUMMA in which {\bf A} or {\bf B} are stationary are also possible; transposed multiplies, {\em e.g.}\ {\bf C} = {\bf A}{\bf B}$^{\dagger}$, are also relatively simple to handle \cite{VanDeGeijn:1997uf,Schatz:2012wy}).
Each pair of broadcasts is followed by a rank-$k$ update, $C_{ij} \gets A_{ik} B_{kj} + C_{ij}$, (Einstein summation convention is used throughout).
Original SUMMA papers by van de Geijn and Watts \cite{VanDeGeijn:1997uf} and by Agarwal {\em et al.}\ \cite{Agarwal:1994hj} considered versions of the algorithm that overlapped computation and communication by pipelining and preemptive broadcasts, respectively, and other broadcast variants have been considered \cite{Schatz:2012wy}, including topology-specific broadcasts \cite{Solomonik:2011jm}.
For simplicity, we present a SUMMA version with preemptive broadcasts in Figure~\ref{alg:basesumma}.

\begin{algorithm}[h]
  \caption{SUMMA with non-blocking broadcast}
  \begin{algorithmic}[1]
    \State \Call{Broadcast}{$A(*,0), 0, row\_group$}
    \State \Call{Broadcast}{$B(0,*), 0, col\_group$}
    \For{$k = 0, \dots, K-1$}
      \If {$k+1 < K$}
        \State $row\_root \gets (k+1) \bmod cols$
        \State $col\_root \gets (k+1) \bmod rows$
        \State \Call{Broadcast}{$A(*,k+1), row\_root, row\_group$}
        \State \Call{Broadcast}{$B(k+1,*), col\_root, col\_group$}
      \EndIf
      \State \Call{Wait}{$A(*,k)$}
      \State \Call{Wait}{$B(k,*)$}
      \State $C(*,*) \gets \alpha A(*,k) \cdot B(k,*) + C(*,*)$
    \EndFor
  \end{algorithmic}
  \label{alg:basesumma}
\end{algorithm}

DIMMA, an early variation of SUMMA introduced by Choi in 1998, improved
performance of synchronous SUMMA by realizing that the order of broadcasts in
the original algorithm coupled with communication barriers created significant
slack in communication \cite{Choi:1997cp,Choi:1998vw}.
Iterations were reordered in DIMMA such that each node broadcasts all of its data in succession as opposed to the round-robin approach in SUMMA. Similar improvements can be attained by overlapping communication and computation \cite{Agarwal:1994hj}.

SUMMA was recently extended to sparse MM (SpSUMMA) by Bulu\c{c} and Gilbert \cite{Buluc:2008vn,Buluc:2012bp}.
This algorithm is similar to dense SUMMA, except matrix sub-blocks are stored in a doubly compressed sparse column (DCSC) format and sparse generalized matrix-multiplication (SpGEMM) is used to compute rank-$k$ updates.
The main challenges of all sparse MM algorithms, including SpSUMMA, are the increased relative costs of communication compared to the dense case, load imbalance, and the relatively low intra-node performance of sparse matrix kernels.

The problem of load imbalance does not appear in dense SUMMA implementations as the work load is nearly-optimally balanced.  With random sparsity, approximate load balance is achieved in the asymptotic limit, however with structured sparsity ({\em e.g.} matrices with decay) one does not expect rows/columns to be uniformly filled.

In the regime of high sparsity (low matrix fill-in factors) the communication time dominates the computation time due to the effectively-reduced benefit of blocking.
The increasing role of communication in sparse MM can be somewhat alleviated by communication hiding. Bulu\c{c} and Gilbert discussed potential benefits of communication
hiding for spa-rse MM in Ref. \cite{Buluc:2008vn} but did not pursue this approach in their experiments due to lack of quality one-sided communication tools \cite{Buluc:2012bp}.
Another possibility is to minimize communication, {\em e.g.}\ by switching to 3D MM \cite{Ballard:2013dy}.
Although sparse 2D SUMMA is not strongly scalable, nevertheless good scalability of SpSUMMA was demonstrated in practice \cite{Buluc:2012bp}.


\section{Task-based SUMMA Formulation}

Our work to improve SUMMA is motivated by the needs of computation on matrices/tensors with irregular
low-rank structure of their blocks. Some of the challenges of computing with such data are similar to the general challenges of sparse MM: increased communication/computation ratio and lack of load balance. The latter is compounded by the desire to use physics-based blocking of matrix dimensions as well as non-standard data representations. To address these challenges, we set to investigate a task-based formulation of SUMMA algorithm designed to partially offset some communication latency and to alleviate the load imbalance.
Prior efforts to reformulate dense matrix multiplication using a tasks-based programming models
are known \cite{pdsec,baruch:2001}; the novelty of our effort is the focus on dense matrices/tensors with irregular structure ({\em i.e.}\ block-rank-sparsity) that cannot be handled straightforwardly using the standard dense-only approaches.

In this section we briefly describe the design of our algorithm, by highlighting the differences with the procedural SUMMA implementations\cite{VanDeGeijn:1997uf}.
We first analyze the data dependencies of discrete operations in the procedural SUMMA implementation.
We then discuss the task composition and dependencies of our modified implementation.
For simplicity, we only consider the 2D SUMMA implementation; though our approach is applicable to 3D and 2.5D variant of SUMMA.

\subsection{Control Flow of Standard SUMMA}
\label{sec:ProceduralSUMMADependencyAnalysis}

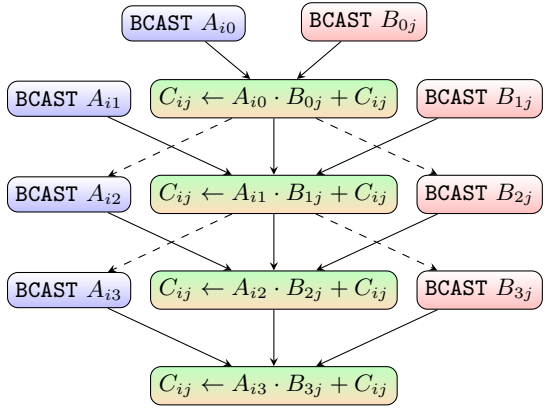
\begin{figure}[h!]
  \centering
  \input{standard_summa.tikz}
  \caption{A directed acyclic graph of the procedural SUMMA implementation consisting of broadcast ({\tt BCAST}) and rank-$k$ updates. Solid edges indicate data dependencies, and dashed edges indicate sequence dependencies.}
  \label{dag:naivesumma}
\end{figure}

Like all dense MM algorithms, SUMMA consists of tightly synchronized data movement and computation (see Algorithm~\ref{alg:basesumma}).
Namely, the rank-$k$ update, $C_{ij} \gets A_{ik} B_{kj} + C_{ij}$, of each iteration depends on the data from the broadcasts of the corresponding panels of {\bf A} and {\bf B} as well as the previous iteration's rank-$k$ update. In addition to these {\em data dependencies}, each broadcast is synchronized with a prior rank-$k$ update since communication operations are initiated at the beginning of each SUMMA iteration, as shown in Figure~\ref{alg:basesumma} (we denote such {\em sequence dependencies} by dashed lines).
Such a design ensures that only a minimal memory overhead occurs (although technically, nonblocking broadcasts require more memory than optimal).
However, this design also limits the work available to each processor (see the  Figure~\ref{dag:naivesumma}), and therefore the amount of parallelism.
Specifically, we can parallelize the rank-$k$ update of $C$ as well as the column and row broadcasts of $A$ and $B$, but SUMMA iterations --- although almost independent from one another --- are executed serially, one after the other. 
Furthermore, such design is not tolerant of any source of load imbalance, due to, for example, processor clock variation, network congestion, slack in communication, or --- most important for us --- due to the inhomogeneity of data from block size or rank variation.

\subsection {Task-Based Multiple-Issue SUMMA}

To tolerate the data inhomogeneity, whether due to block sparsity or block-rank sparsity 
and address the above deficiencies of the standard 2D SUMMA, we introduced the following modifications:
(1) overdecomposition of data and work,
(2) elimination of dependencies between SUMMA iterations,
(3) scheduling of multiple iterations of SUMMA at once, and
(4) the use of task-decomposed broadcasts.

Although the standard dense SUMMA allows near-perfect load balance of data by
using cyclic embedding of matrices onto 2D process grid, it is not feasible to 
maintain such load balance with irregular data structures throughout the MM computation, even with uniform blocking.
In fact the dimension blocking is a central concept in the {\sc TiledArray} library since it arises naturally in physical problems (hence the need to support arbitrary block sizes, including nonuniform blocking). Thus we support an  arbitrary blocking to match the physics of the problem, but assume that
the data is overdecomposed (i.e. there are many blocks per processor); this helps to improve load balance of the data.

To improve the load balance of work, the artificial dependencies between SUMMA iterations are eliminated and
multiple SUMMA iterations are scheduled at once.
This increases the average number of computational tasks per processor, and thus improves the load balance
under the assumption that the amount of work performed by each task is randomly distributed.
Note that the dependencies between iterations is the result of GEMM operations (${\bf C} = {\bf A} {\bf B} + {\bf C}$) via updates of the result matrix, ${\bf C}$.
To decouple this data dependency, we assume that there is
enough memory available to split the rank-$k$ update operation into two separate tasks: a matrix-multiplication task producing a temporary block, $C_{ij}^{(k)} = A_{ik} \cdot B_{kj}$, and a reduction of the temporary into the result, $C_{ij} = C_{ij}^{(k)} + C_{ij}$.
In our implementation, we mitigate the additional memory cost by automatically switching between GEMM and split matrix-multiply-and-reduces updates based on the data availability.
Specifically, the GEMM update is used when only a single thread requires access to a sub-block and the split matrix-multiply and reduces tasks are used when two or more threads must update the same sub-block.
This update scheme is essential for high computational throughput and resource management in our task-based SUMMA formulation.

Scheduling multiple SUMMA iterations at once impacts both performance and memory consumption.
In that sense the multiple-issue SUMMA is similar to 2.5D and 3D MM algorithms \cite{Solomonik:2011bra}
that trade off memory to increase concurrency.
The number of SUMMA iterations scheduled concurrently is a user-adjustable parameter and can be tuned
to minimize memory consumption (see next Section) or to maximize the computational throughput.
In practice, we found that the throughput is maximized
when each process initiates $I_{\rm opt}$ SUMMA iterations, where for a process grid with $P_{\rm r}$ rows and $P_{\rm c}$ columns $I_{\rm opt} = \max(2, \min(P_{\rm r},P_{\rm c}))$; this is the default schedule depth, unless specified otherwise.
As scheduled SUMMA iterations are retired, additional iterations are scheduled, in a pipelined fashion.

Lastly, data broadcasts are decomposed into several smaller tasks, which allows overlap of communication and computation at the task level.
For example, a given process can start the computational work for a given iteration as soon as the minimal amount of data is available, and execute concurrently with the remaining communication tasks.
Integration of communication operations allows more efficient work scheduling based on the availability of data in our task approach and improves tolerance of irregular, high-latency communication.

\subsection{Memory Overhead of Multiple-Issue SUMMA \label{sec:tbsummamemory}}

The dynamic scheduling of computation and communication trades off predictable bounds on resource use, in particular memory, for high performance. The maximum memory requirement per process of our multiple-issue SUMMA, for the multiplication of $\mathbb{R}^{M \times K}$ and $\mathbb{R}^{K \times N}$ dense matrices with average block sizes of $m \times k$ and $k \times n$, respectively, is proportional to:
\begin{equation}
  I \left( \frac{Mk}{P_{\rm r}} + \frac{Nk}{P_{\rm c}} \right) + \frac{MN + MK + KN}{P_{\rm r} P_{\rm c}}
  \label{eqn:task_summa_max_mem}
\end{equation}
where $I$ is the number of concurrently-scheduled iterations; and $P_{\rm r}$ and $P_{\rm c}$ are the number of rows and columns in the process grid, respectively.
Eq. \eqref{eqn:task_summa_max_mem} correspond to the memory requirements of the 2D, bulk-synchronous SUMMA when $I=1$.
Our implementation also uses additional temporary storage for replicated sub-blocks of the result matrix (not shown in Eq. \eqref{eqn:task_summa_max_mem}), but this is a negligible overhead in large problems where the number of blocks is much larger than the number of cores.
If, for simplicity, we assume square matrices and process grid ($M = N = K$, $m = n = k$, $P_{\rm r} = P_{\rm c} = \sqrt{P}$) then Eq. \eqref{eqn:task_summa_max_mem} reduces to
\begin{equation}
  I \frac{2Nk}{\sqrt{P}} + \frac{3N^2}{P}
  \label{eqn:task_summa_max_mem_sq}
\end{equation}
where  $I (2Nk/\sqrt{P})$ is the memory overhead due to the partial replication of the argument matrices.
Under the assumption of overdecomposition, the block size $k$ is much smaller than $N / \sqrt{P}$,
hence the first term in Eq. \eqref{eqn:task_summa_max_mem_sq} is much smaller than the second, even for $I>1$.
Note that Eq. \eqref{eqn:task_summa_max_mem} is derived under the pessimistic assumption that the rate of computation
is much lower than the rate of communication. In practice, however, as soon as the data arrives it is consumed by the compute tasks, hence the effective value of $I$ is lower than the actual number of SUMMA iterations
scheduled.

The average memory overhead for block-sparse SUMMA can be estimated by scaling each matrix, and corresponding replicated blocks, by the fraction of non-zero elements.
In addition, some process may not contain non-zero data for a given iteration.
Therefore, the memory overhead of the replicated blocks by $\langle P_{\rm r} \rangle / P_{\rm r}$ and $\langle P_{\rm c} \rangle / P_{\rm c}$.
With these modifications, Eq. \eqref{eqn:task_summa_max_mem} becomes:
\begin{equation}
  \begin{split}
  I \left( (1 - z_{\mathbf A}) \frac{\langle P_{\rm c} \rangle}{P_{\rm c}} \frac{M k}{P_{\rm r}} +
  (1 - z_{\mathbf B}) \frac{\langle P_{\rm r} \rangle}{P_{\rm r}} \frac{N k}{P_{\rm c}} \right) + \\
  \frac{(1 - z_{\mathbf C}) MN + (1 - z_{\mathbf A}) MK + (1 - z_{\mathbf B}) KN}{P_{\rm r} P_{\rm c}}
  \end{split}
  \label{eqn:sparse_summa_max_mem}
\end{equation}
where $1-z_{\mathbf X}$ the fraction of non-zero elements for matrix ${\mathbf X}$, and $\langle P_{\rm r} \rangle$ and $\langle P_{\rm c} \rangle$ are the expectation values for the number of row and column processes, respectively, with non-zero contributions in a single SUMMA iteration.

Note that our implementation may also use less memory than that given in Eqs. \eqref{eqn:task_summa_max_mem} and \eqref{eqn:sparse_summa_max_mem}. 
By overdecomposing the rank-$k$-update and broadcast tasks such that each matrix-multiplication task only computes a small sub-block of the local result matrix.
This decomposition is similar to the decomposition of work between nodes within a SUMMA iteration, but without the analogous communication (dependencies) between computation tasks.
It allows streaming of data so that processes do not need to hold all data for the block row and column in a given iteration.
In addition, higher priority is given to tasks that free resources, which limits the accumulation of temporary storage space.
Unfortunately, {\em a priori} prediction of the actual memory usage is not
possible due to the non-deterministic order of execution.

With block-rank-sparse matrices, used in our square root inverse algorithm (see Sections \ref{sec:results-low-rank-rep} and \ref{sec:results-sqrtinv} for details), the exact memory requirements are similarly non-deterministic as the rank may grow or shrink based on the order of operations and nature of the input data.
However, the upper memory bound is equal to that of the full-rank (dense) computation.
Unfortunately, the rank of each block cannot easily be determined until runtime, making {\em a priori} estimations of the memory requirements difficult. 

\section{Results} \label{sec:result-intro}


In this section, we: (a) discuss the implementation details of tensor arithmetic in low-rank form,
(b) evaluate the performance of our methods with this low-rank data structure by computing the
inverse square root of matrices used in the quantum chemistry domain, and
(c) demonstrate the performance of our implementation for dense matrix multiplication.

Our 2D task-based SUMMA algorithm is implemented in
the {\sc TiledArray} (TA) library \cite{TiledArray}, which uses the parallel runtime of
{\sc MADNESS} \cite{MADNESS} to manage the low-level details of task scheduling
and data movement. Unless noted otherwise, we used TA 0.5.0-alpha, and {\sc MADNESS} dated 08/26/2015.
Serial DGEMM from the Intel Math Kernel Library (MKL) 11.2.3
was used as the block multiply-add kernel in our MM tasks. TA and {\sc MADNESS} were
compiled with the Intel Parallel Studio XE 15.3 and Intel MPI 5.0. Both TA and MADNESS
can be obtained under the terms of the GNU General Public License.

These tests were performed on a 408-node Cray CS-300 cluster, with two eight-core Intel
Xeon E5-2670 CPUs and 64 GB of memory per compute node. In addition, we performed
dense MM scaling tests on the IBM BlueGene/Q {\em Mira} supercomputer at
Argonne National Laboratory.

\subsection{Clustered Low-Rank Representation} \label{sec:results-low-rank-rep}

\begin{figure*}
\centering
  \begin{minipage}{\columnwidth}
    \centering
    \includegraphics[width=0.9\columnwidth]{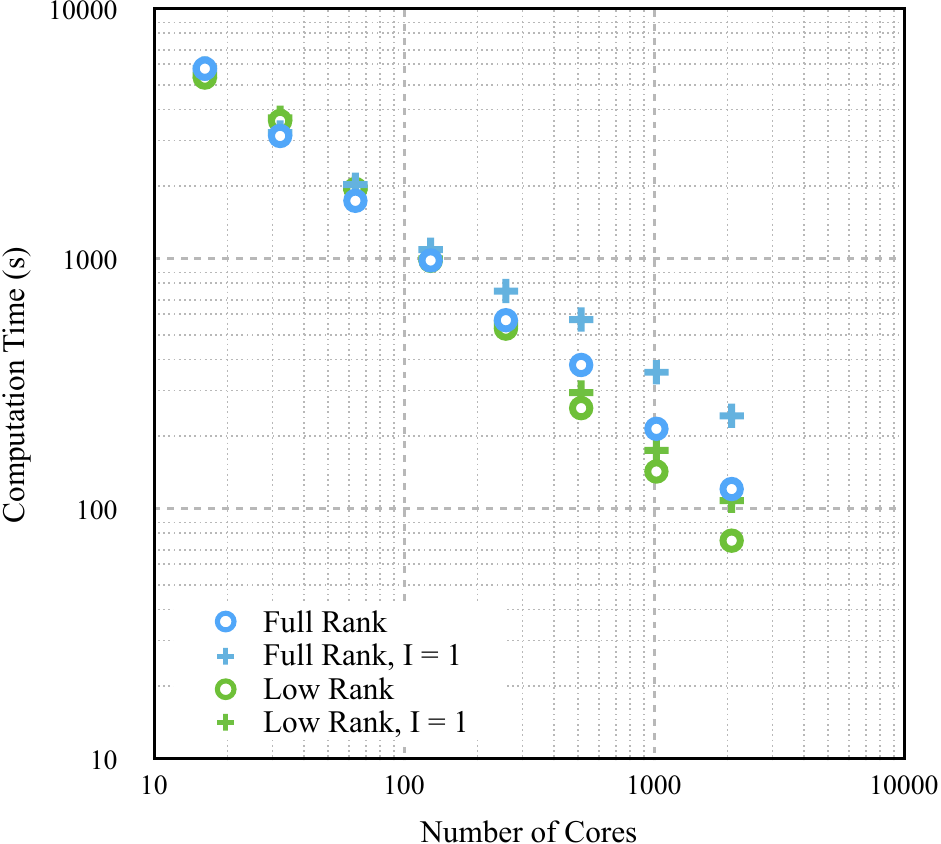}
    \caption{
        Wall time for computing the square root inverse
        of the Coulomb matrix. $I=1$ refers to data obtained with
        the standard single-issue SUMMA; the rest of data obtained with the
        multiple-issue SUMMA.
    }
    \label{fig:sqrt_inv_coulomb}
  \end{minipage}~~~~~~~~%
  \begin{minipage}{\columnwidth}
    \centering
    \includegraphics[width=0.9\columnwidth]{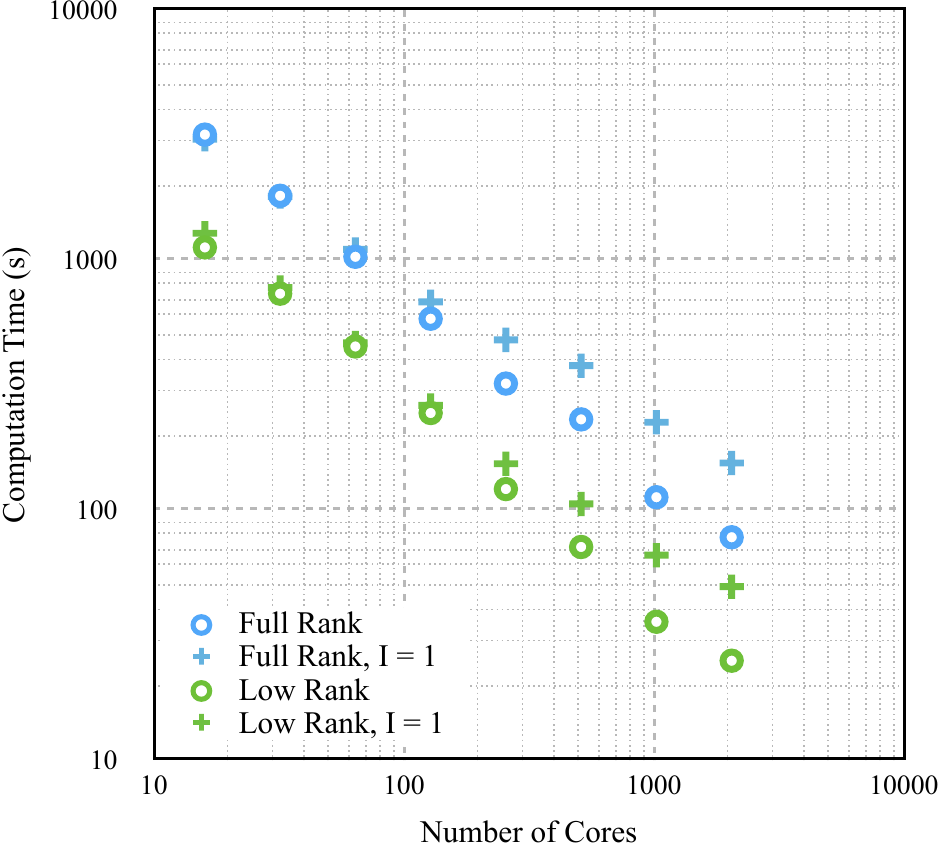}
    \caption{
        Wall time for computing the square root inverse
        of the overlap matrix. $I=1$ refers to data obtained with
        the standard single-issue SUMMA; the rest of data obtained with the
        multiple-issue SUMMA.
    }
    \label{fig:sqrt_inv_overlap}
 \end{minipage}
\end{figure*}

To recover data sparsity in tensors appearing in quantum physics applications
we developed the Clustered Low-Rank (CLR)
representation \cite{LewisCalvinValeevToBeSubmitted} that is a general, hierarchy-free
compressed tensor format. In this representation, each
$\mathbf{M}_{\mathcal{I}\mathcal{J}}$ block of matrix $\mathbf{M}$ is approximated by a low-rank decomposition of the form
$\mathbf{M}_{\mathcal{I} \mathcal{J}} \approx \mathbf{X}
\mathbf{W}^{\dagger}$, where for a given $\mathbf{M}_{\mathcal{I} \mathcal{J}} \in \mathbb{R}^{m \times n}$ of rank $r$, $\mathbf{X}$ is $m \times r$ and $\mathbf{W}$ is $n \times r$. $\mathbf{X}$ and $\mathbf{W}$ were constructed from a rank-revealing QR decomposition,\cite{QuintanaOrti:2006df}
$\mathbf{M}_{\mathcal{I} \mathcal{J}}\mathbf{P} = \mathbf{Q} \mathbf{R}$. Thus $\mathbf{X} \equiv \mathbf{Q}_{r}$ and $\mathbf{W}^{\dagger} \equiv \mathbf{R}_{r} \mathbf{P}^{\dagger}$ where subscript $r$ indicates that only $r$
most significant columns and rows of $\mathbf{Q}$ and $\mathbf{R}$ were kept; $r$ was determined so that the Frobenius norm of the low-rank approximation error, $\|\mathbf{M}_{\mathcal{I} \mathcal{J}} - \mathbf{X} \mathbf{W}^{\dagger} \|_{\rm F}$ is less than or equal to $\epsilon_{\rm lr}$, a user-defined threshold. For blocks where the rank of
$\mathbf{M}_{\mathcal{I} \mathcal{J}}$ is greater than half of full rank, the
blocks are stored in their full representation.

To maintain compression, all block operations, such as addition and
GEMM, are performed in low-rank form directly whenever possible.  Low-rank matrix multiplication
of $\mathbf{C} = \mathbf{A} \mathbf{B}$, where all matrices appear in low-rank
form, uses the following steps.  First, create a temporary $\mathbf{Z} =
(\mathbf{W}^{\mathbf{A}})^{\dagger}\mathbf{X}^{\mathbf{B}}$, where the superscript letters designate the matrix that $\mathbf{X}$ and $\mathbf{W}$ were constructed from, leaving
\begin{align}
    \mathbf{X}^{\mathbf{C}} (\mathbf{W}^{\mathbf{C}})^{\dagger} &= 
        \mathbf{X}^{\mathbf{A}} \mathbf{Z} 
        (\mathbf{W}^{\mathbf{B}})^{\dagger} \text{.}
\end{align}
Then contract $\mathbf{Z}$ with $\mathbf{X}^{\mathbf{A}}$ or
$\mathbf{W}^{\mathbf{B}}$ such that the output rank is minimized; contracting $\mathbf{Z}$
with $\mathbf{X}^{\mathbf{A}}$ yields 
$\mathbf{X}^{\mathbf{C}} = \mathbf{X}^{\mathbf{A}} \mathbf{Z}$ and
$\mathbf{W}^{\mathbf{C}} = \mathbf{W}^{\mathbf{B}}$. 

Low-rank addition of $\mathbf{C} = \mathbf{A} + \mathbf{B}$ may be
viewed as a sum of outer products:
\begin{align} \label{eq:lr_addition}
    \mathbf{X}^{\mathbf{C}} (\mathbf{W}^{\mathbf{C}})^{\dagger} &= 
        \sum_{i = 1}^{r_A} \mathbf{x}_i^{\mathbf{A}} * (\mathbf{w}^{\mathbf{A}}_i)^{\dagger}
        + \sum_{j = 1}^{r_B} \mathbf{x}_j^{\mathbf{B}} * (\mathbf{w}^{\mathbf{B}}_j)^{\dagger} \text{,}
\end{align}
where $\mathbf{x}_i^{\mathbf{A}}$, $\mathbf{w}^{\mathbf{A}}_i$,
$\mathbf{x}_j^{\mathbf{B}}$, and $\mathbf{w}^{\mathbf{B}}_j$ represent column
vectors of the corresponding matrix and $r_A$ and $r_B$ are the ranks of $\mathbf{A}$ and $\mathbf{B}$ respectively.  From Eq. \eqref{eq:lr_addition}, it is
clear that the low-rank
matrices representing $\mathbf{C}$ are the union of the input matrices, with
$\mathbf{X}^{\mathbf{C}} = \{ \mathbf{X}^{\mathbf{A}}
\mathbf{X}^{\mathbf{B}}\}$ and $\mathbf{W}^{\mathbf{C}} = \{
\mathbf{W}^{\mathbf{A}} \mathbf{W}^{\mathbf{B}}\}$. 

Finally, in order to compress a matrix that
is already in a low-rank form $\mathbf{C} = \mathbf{X} (\mathbf{W})^{\dagger}$
to a more compressed form $\mathbf{C} = \mathbf{X'} (\mathbf{W'})^{\dagger}$ we
first perform a QR decomposition of $\mathbf{X}$ and $\mathbf{W}$ 
\begin{align}
    \mathbf{X}' (\mathbf{W}')^{\dagger} &=
        \mathbf{Q}^{\mathbf{X}} \mathbf{R}^{\mathbf{X}} (\mathbf{R}^{\mathbf{W}})^{\dagger} (\mathbf{Q}^{\mathbf{W}})^{\dagger}
\end{align}
and then form a temporary matrix $\mathbf{M} = \mathbf{R}^{\mathbf{X}}
(\mathbf{R}^{\mathbf{W}})^{\dagger}$ giving us
\begin{align}
    \mathbf{X}'(\mathbf{W}')^{\dagger} &=
        \mathbf{Q}^{\mathbf{X}} \mathbf{M} (\mathbf{Q}^{\mathbf{W}})^{\dagger} \text{.}
\end{align}
We can perform a low-rank decomposition of $\mathbf{M} =
\mathbf{X}^{\mathbf{M}} (\mathbf{W}^{\mathbf{M}})^{\dagger}$ creating a new
compressed representation of the original block where 
$\mathbf{X}' = \mathbf{Q}^{\mathbf{X}} \mathbf{X}^{\mathbf{M}}$ and
$\mathbf{W'} = \mathbf{Q}^{\mathbf{W}} \mathbf{W}^{\mathbf{M}} $. After compression, the rank
of $\mathbf{C}$ has been decreased from $r_A + r_B$ to the rank of
$\mathbf{M}$.

Arithmetic on CLR matrices is implemented in terms of the above low-rank block arithmetic. To avoid computing
blocks whose norm will be smaller than target precision, block $\mathbf{C}_{\mathcal{IJ}}$ of result matrix $\mathbf{C}$ is only computed
if its Frobenius norm {\em estimate} satisfies $||\mathbf{C}_{\mathcal{IJ}}||_{F} \leq \epsilon_{\rm sp} \,\mathrm{area}(\mathbf{C}_{\mathcal{IJ}})$, where $\mathrm{area}(\mathbf{C}_{\mathcal{IJ}})$ is the number of elements in  $\mathbf{C}_{\mathcal{IJ}}$, and $\epsilon_{\rm sp}$ is a user-defined parameter. Estimated of Frobenius norms of sums and products of blocks are estimated using the upper bounds provided by the triangle inequality and submultiplicativity of the Frobenius norm, e.g. $||\mathbf{A}_{\mathcal{IJ}} + \mathbf{B}_{\mathcal{IJ}}||_{F} \leq ||\mathbf{A}_{\mathcal{IJ}}||_{F} + ||\mathbf{B}_{\mathcal{IJ}}||_{F}$, hence the right-hand side of the inequality is used as the estimate. Complete details of screening the arithmetic operations are provided in Ref. \cite{LewisCalvinValeevToBeSubmitted}.

To summarize: the accuracy of CLR representation and arithmetic is controlled by two user-defined parameters, $\epsilon_{\rm lr}$ and $\epsilon_{\rm sp}$. As $\epsilon_{\rm lr} \to 0$, the CLR representation becomes exact; similarly, when $\epsilon_{\rm sp} \to 0$ the arithmetic on CLR matrices becomes exact.


\subsection{Iterative Square Root Inverse} \label{sec:results-sqrtinv}

To compute the inverse square root of a matrix $\mathbf{M}$, where the matrix square root is given by the matrix $\mathbf{M}^{1/2}$ such that $\mathbf{M}^{1/2} \mathbf{M}^{1/2} = \mathbf{M}$ and $\mathbf{M}^{-1/2}\mathbf{M}^{1/2} = \mathbf{I}$, we used the
iterative matrix-mulitplication approach in \cite{Jansik:2007dp} based on the Newton-Schulz method
\cite{Higham:1997gz,Niklasson:2004in} consisting of the following steps in each
iteration:
\begin{align}
    \mathbf{X}_{n} &= \alpha \mathbf{Y}_n \mathbf{Z}_n, \\
    \mathbf{T}_{n} &= \frac{1}{8}\left(-10 \mathbf{X}_n + 
        3\mathbf{X}_n^2 + 15 \mathbf{I} \right), \\
    \mathbf{Z}_{n+1} &= \mathbf{Z}_n \mathbf{T}_n, \\
    \mathbf{Y}_{n+1} &= \mathbf{T}_n \mathbf{Y}_n, 
\end{align}
where $\mathbf{I}$ is the identity matrix, and upon convergence $\sqrt{\alpha} \mathbf{Z} = \mathbf{M}^{-1/2}$ is the sought square root inverse. The starting guesses are $\mathbf{Z}_0 = \mathbf{I}$, $\mathbf{Y}_0 = \mathbf{M}$, and
$\alpha$ is chosen to scale $\mathbf{M}$ such that $\|\alpha \mathbf{M} - \mathbf{I}\|_2
\leq 1$.


Square root inverses of the overlap and Coulomb operator matrices, $\mathbf{S}$ and $\mathbf{V}$ respectively \cite{SzaboOstlund}, were computed for a cluster of 190 water molecules with the cc-pVTZ-RI Gaussian basis, consisting of 141 basis functions {\em per} water molecule, for a total basis size of \num[group-separator={,}]{26790} basis functions.
The matrices use natural blocking, i.e. each 141 by 141 block spans basis functions associated with the corresponding water molecule.
$\epsilon_{\text{sp}} = 10^{-13}$ and $\epsilon_{\text{lr}} = 10^{-6}$ were used.

Our performance tests executed 10 iterations of this algorithm;
the resulting wall times are reported in Figs. \ref{fig:sqrt_inv_coulomb} and \ref{fig:sqrt_inv_overlap}.
Good strong scaling is observed: for example, for the Coulomb matrix in dense and low-rank representations the parallel efficiency is maintained at 43\% and 59\%, respectively, upon scaling from 1 to 128 nodes.
For the overlap matrix, the corresponding figures are 28\% and 31\%. The observed differences in scaling between
overlap and Coulomb are likely due to the much faster (exponential) decay of overlap matrix with distance and hence greatly reduced amount of work due to more efficient screening of matrix operations: without such screening the time to compute the inverses
of Coulomb and overlap matrices in full-rank representation would take exactly the same amount of time, whereas in practice
the inversion of the overlap is roughly a factor of 2 faster than that of the Coulomb matrix in full-rank form.

Unlike $\mathbf{S}$, $\mathbf{V}$ does not have any block-sparsity, only block-rank-sparsity. Hence computing $\mathbf{V}^{-1/2}$ in the full-rank form essentially performs the same amount of work as would an implementation
using a standard dense matrix package. The apparent performance of the full-rank inverse was estimated by counting FLOPs from 4 GEMMs per each iteration; the resulting performance on 1 node for the Coulomb case is 263 GLOPs, or $79\%$ of theoretical peak, which is close to optimal. On 128 nodes the apparent throughput is respectable 99 GLOPs.

The benefit of the low-rank representation is immediately apparent for the overlap matrix on 1 node, whereas for the Coulomb matrix
the low-rank implementation outperforms the full-rank counterparts when the number of processors is large, due to the greatly reduced
amount of communication in the low-rank case. Also obvious is the importance of scheduling multiple SUMMA iterations for performance
in the high-end strong scaling regime for both dense and low-rank matrices. For example, on 128 nodes the use of single-issue SUMMA when computing $\mathbf{S}^{-1/2}$ in low-rank form is 97\% slower than multiple-issue SUMMA.


\begin{figure}[t]
  \centering  
  \includegraphics[width=0.9\columnwidth]{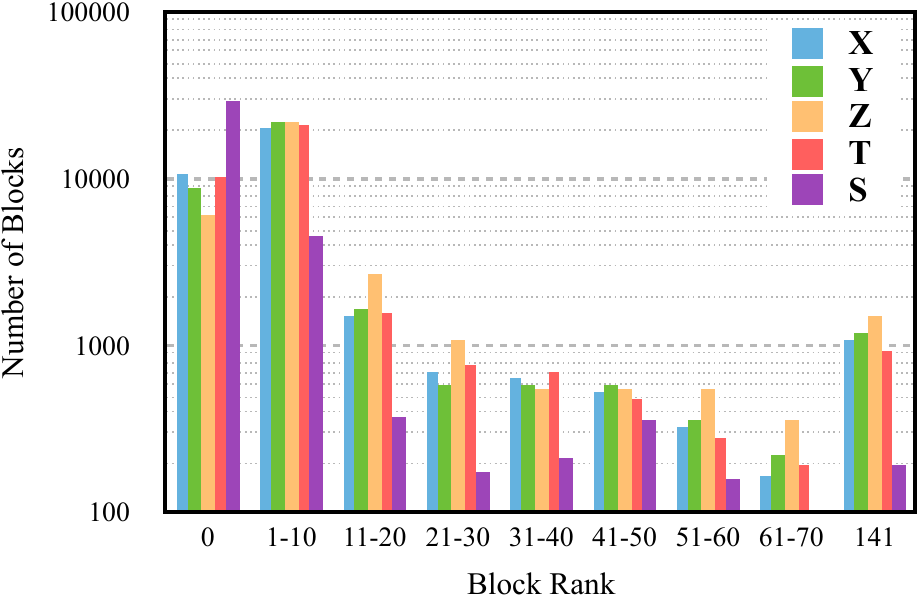}
  \caption{
      Distribution of block ranks of the intermediate matrices from the 5th iteration of the inverse
      square root computation of the overlap matrix.
      A rank of ``0'' indicates a block contains no data, and ``141'' indicates full rank.
  }
  \label{fig:rank_dist}
\end{figure}

Figure \ref{fig:rank_dist} demonstrates the extent of the data inhomogeneity in the matrices
involved in these computation. It shows a histogram of block ranks of 5th iteration intermediates $\mathbf{X}$,
$\mathbf{Y}$, $\mathbf{Z}$, $\mathbf{T}$, as well as the input $\mathbf{S}$ matrix. Most blocks in these matrices
have very low ranks. However, a significant number of blocks have close-to-full rank, which is typical behavior for matrices with decay.

\subsection{Dense Matrix Multiplication} \label{sec:densemm_result}

In the limit where the low-rank threshold and sparsity thresholds become zero, {\em i.e.}\ $\epsilon_{\rm lr} \to 0$ and $\epsilon_{\rm sp} \to 0$, CLR matrices become full-rank (dense) matrices.
In addition, the features of our approach that make it appropriate for rank-structured matrix multiplication also lead to excellent performance for dense matrices.
Thus, in this section we evaluate the performance of our SUMMA formulation in {\sc TiledArray} (TA) \cite{TiledArray} for the case of dense, full-rank matrices in absolute terms and relative to that of {\sc ScaLAPACK} \cite{Blackford:1997ta} and {\sc Cyclops Tensor Framework} (CTF) 1.1 \cite{Solomonik:2012vw,Solomonik:2012vg}, a state-of-the-art, dense tensor contraction implementation based on 2.5D SUMMA.
Performance was evaluated in two prototypical setups: on a commodity cluster and on a high-end IBM BlueGene/Q {\em Mira} supercomputer at Argonne National Laboratory.

On the commodity cluster, we used the version of ScaLAPACK provided by Intel MKL 11.2.3. 
CTF was compiled with the same Intel compiler described at the beginning of this section. 
Unlike TA, however, ScaLAPACK and CTF use the threaded version of BLAS DGEMM routine from MKL for their multiply-add kernels.

Due to the constraints of the existing parallel dense MM software and to simplify the performance analysis, we perform multiplications of square, uniformly-blocked, double-precision matrices; our implementation is completely general.
The matrix size used in the tests on the commodity cluster was \num[group-separator={,}]{32768}, with block size of $256$ used for {\sc TiledArray}.
Reported wall times are averages of 15 repeated multiplications of same input matrices.

\begin{figure}[t]
  \centering  
  \includegraphics[width=0.9\columnwidth]{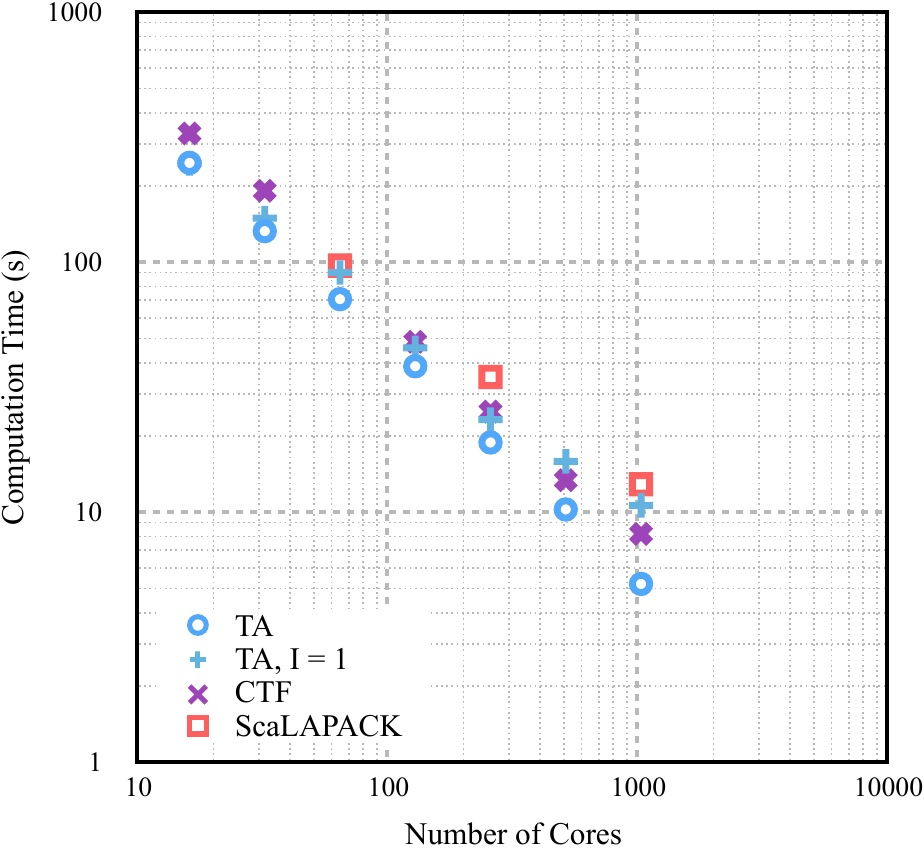}
  \caption{Wall time of square dense matrix multiplication with {\sc TiledArray}, {\sc CTF}, and {\sc ScaLAPACK}. Matrix size = \num[group-separator={,}]{32768}.}
  \label{fig:dense_mm}
\end{figure}

In Figure~\ref{fig:dense_mm}, we show the result of our strong scaling tests, which vary from 16 to 1024 cores.
Each of the MM software packages shows linear scaling across the range of the tests.
However, of the three packages, TA achieved the lowest computational time in all cases.
In fact, the performance differences between TA, ScaLAPACK, and CTF are significant.
On 64 compute nodes, ScaLAPACK took between 1.36 $\sim$ 2.50 times longer to complete relative to TA, while CTF took 1.25 $\sim$ 1.59 times longer to complete.
The parallel efficiency of TA MM relative to 1 node is between 84.3\% and 63.7\% on 64 nodes.

We also studied the performance of TA with the number of concurrent iterations, $I$, set to one to demonstrate the performance benefits of multiple-issue SUMMA iterations.
As expected, the performance of TA with $I = 1$ was approximately the same on a single node, but performs much more slowly with larger node counts.
Scaling with a single-issue (standard) SUMMA is linear, but substantially slower than that of multiple-issue SUMMA, and is consistent with the square root inverse performance tests in Section~\ref{sec:results-sqrtinv}.

\begin{figure}[h!]
  \includegraphics[width=0.9\columnwidth]{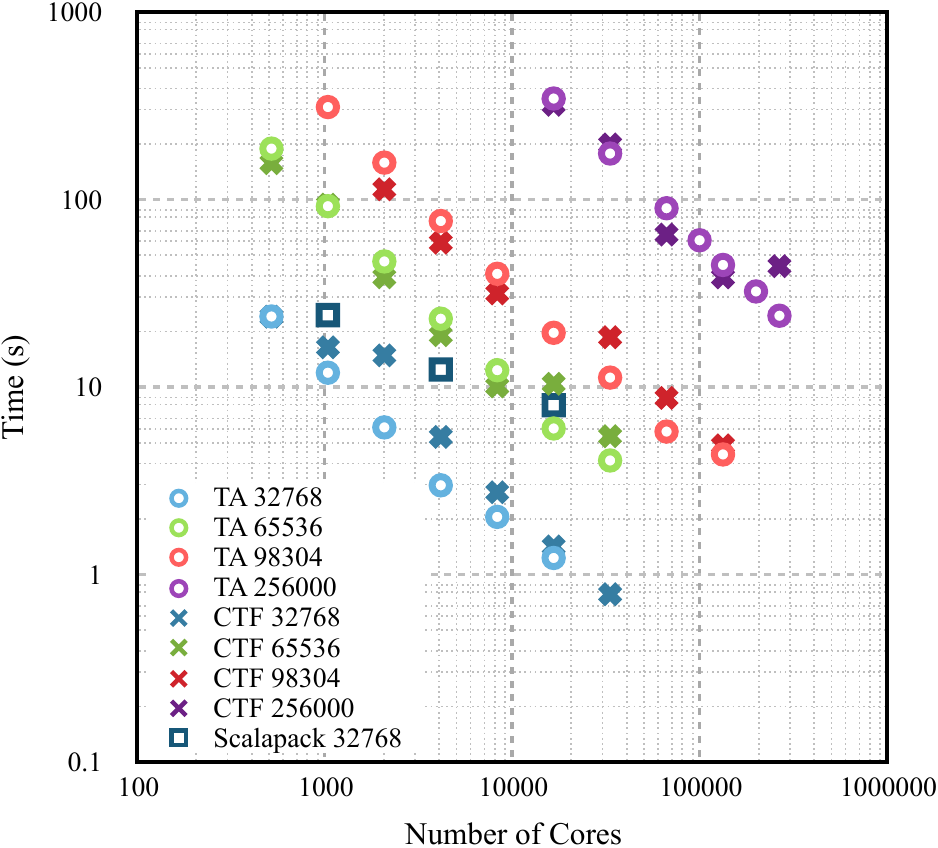}
  \caption{Wall time of square dense matrix multiplication with {\sc TiledArray}, {\sc CTF}, and {\sc ScaLAPACK} on IBM BlueGene/Q. The matrix sizes included in these tests are \num[group-separator={,}]{32768}, \num[group-separator={,}]{65536}, \num[group-separator={,}]{98304}, and \num[group-separator={,}]{256000}.}
    \label{fig:bgqtime}
\end{figure}

Finally, we evaluated the performance of our SUMMA implementation on the IBM BlueGene/Q {\em Mira} supercomputer at Argonne National Laboratory.
In these performance tests, we used TA 0.4.0-alpha, {\sc MADNESS} dated 03/02/2015, {\sc ScaLAPACK} 2.0.2, and CTF 1.1.
The parallel version of ESSL 5.1 was used for the CTF and ScaLAPACK performance tests, while the serial version is used for TA.
TA and {\sc MADNESS} were compiled with GCC 4.7.2; CTF is compiled with XLC 12.1.
Both used the V1R2M2 compiler driver.

In Figure~\ref{fig:bgqtime}, we show the performance of TA, CTF, and {\sc ScaLAPACK} with matrices of various sizes. {\sc TiledArray} used $128\times128$ blocks. The number of cores varies from 512 to 262144.
Both TA and CTF are found to be  competitive with each other, with excellent strong and weak scaling, though each behaves differently across the range of the tests.
We found {\sc ScaLAPACK}, on the other hand, to scale very poorly relative to the TA and CTF, with almost an order of magnitude difference in computational time on 16384 cores.

\section{Conclusions}

We presented a task-based multiple-issue formulation of the Scalable Universal Matrix Multiplication Algorithm (SU-MMA) that performs well for multiplication of matrices with irregular structure, such as block-sparsity and block-rank-sparsity.
An implementation of our algorithm in an open-source library {\sc TiledArray} was used to compute iteratively square-root inverses of two realistic matrices from the domain of quantum chemistry, using block-sparse and block-rank-sparse representations. 
Excellent strong scaling from 16 to 2048 cores was observed on a commodity cluster.
For dense matrices our implementation is competitive with state-of-the-art dense tensor algebra libraries like {\sc Cyclops Tensor Framework}, both on a commodity cluster using up to 1024 cores as well as on a high-end IBM BG/Q supercomputer using up to 262 thousand cores.

\section{Acknowledgments}
This work was supported by the National Science Foundation (awards CHE-1362655, ACI-1047696, and ACI-1450262), Camille Dreyfus Teacher-Scholar Award, and Alfred P. Sloan Fellowship.
We gratefully acknowledge the computer time provided by the Innovative and Novel Computational Impact on Theory and Experiment (INCITE) program. This research used resources of the Argonne Leadership Computing Facility, which is a DOE Office of Science User Facility supported under Contract DE-AC02-06CH11357.
We acknowledge the computer time allocation from Advanced Research Computing at Virginia Tech.

%
\bibliographystyle{hacm}
\bibliography{refs,refs-ev}  
%
%

\end{document}

%% file: standard_summa.tikz
\begin{tikzpicture}[
  ->,
  >=stealth,
  node distance=0.25cm,
  task/.style={rectangle, draw, rounded corners=4pt,top color=white}
]
  
  \node[task,top color=green!25,bottom color=orange!25] (GEMM0) {$C_{ij} \gets A_{i0} \cdot B_{0j} + C_{ij}$};
  \node[task,bottom color=blue!25] (BCASTA0) [node distance=0.5cm,above=of GEMM0,xshift=-1.2cm] {{\tt BCAST} $A_{i0}$};
  \node[task,bottom color=red!25] (BCASTB0) [node distance=0.5cm,above=of GEMM0,xshift=1.2cm] {{\tt BCAST} $B_{0j}$};
  \node[task,top color=green!25,bottom color=orange!25] (GEMM1)   [node distance=0.75cm,below=of GEMM0] {$C_{ij} \gets A_{i1} \cdot B_{1j} + C_{ij}$};
  \node[task,bottom color=blue!25] (BCASTA1) [left=of GEMM0] {{\tt BCAST} $A_{i1}$};
  \node[task,bottom color=red!25] (BCASTB1) [right=of GEMM0] {{\tt BCAST} $B_{1j}$};
  \node[task,top color=green!25,bottom color=orange!25] (GEMM2)   [node distance=0.75cm,below=of GEMM1] {$C_{ij} \gets A_{i2} \cdot B_{2j} + C_{ij}$};
  \node[task,bottom color=blue!25] (BCASTA2) [left=of GEMM1] {{\tt BCAST} $A_{i2}$};
  \node[task,bottom color=red!25] (BCASTB2) [right=of GEMM1] {{\tt BCAST} $B_{2j}$};
  \node[task,top color=green!25,bottom color=orange!25] (GEMM3)   [node distance=0.75cm,below=of GEMM2] {$C_{ij} \gets A_{i3} \cdot B_{3j} + C_{ij}$};
  \node[task,bottom color=blue!25] (BCASTA3) [left=of GEMM2] {{\tt BCAST} $A_{i3}$};
  \node[task,bottom color=red!25] (BCASTB3) [right=of GEMM2] {{\tt BCAST} $B_{3j}$};

  \path (BCASTA0) edge node {} (GEMM0)
        (BCASTB0) edge node {} (GEMM0)
        (GEMM0)   edge node {} (GEMM1)
        (BCASTA1) edge node {} (GEMM1)
        (BCASTB1) edge node {} (GEMM1)
        (GEMM1)   edge node {} (GEMM2)
        (BCASTA2) edge node {} (GEMM2)
        (BCASTB2) edge node {} (GEMM2)
        (GEMM2)   edge node {} (GEMM3)
        (BCASTA3) edge node {} (GEMM3)
        (BCASTB3) edge node {} (GEMM3);

  \path [dashed]
           (GEMM0) edge node {} (BCASTA2)
           (GEMM0) edge node {} (BCASTB2)
           (GEMM1) edge node {} (BCASTA3)
           (GEMM1) edge node {} (BCASTB3);

\end{tikzpicture}